\documentclass[prl,aps,showpacs,twocolumn,amsmath,amssymb]{revtex4}
\usepackage{epsfig,xspace}
\usepackage[pdftitle={Winding angle variance of Fortuin-Kasteleyn contours},
            pdfauthor={Benjamin Wieland and David B. Wilson},
            bookmarks=true,bookmarksopen=true,bookmarksopenlevel=2,
            hyperfootnotes=false]{hyperref}
\begin{document}

\title{Winding Angle Variance of Fortuin-Kasteleyn Contours}

\author{Benjamin Wieland$^*$}
\author{David B. Wilson$^\dag$}
\affiliation{$^*$Department of Mathematics, University of Chicago, Chicago IL 60637, U.S.A.}
\affiliation{$^\dag$Microsoft Research, One Microsoft Way, Redmond WA 98052, U.S.A.}

\begin{abstract}
The variance in the winding number of various random fractal curves,
including the
self-avoiding walk, the loop-erased random walk, contours of FK
clusters, and stochastic Loewner evolution,
have been studied by numerous researchers.  Usually the
focus has been on the winding at the endpoints.  We
measure the variance in winding number at typical points along the
curve.  More generally, we study the winding at points where $k$
strands come together, and some adjacent strands may
be conditioned not to hit each other.  The measured values are
consistent with an interesting conjecture.
\end{abstract}


\maketitle
\newcommand{\sect}[1]{}

%
Duplantier and Saleur \cite{DS} studied the winding angle between the
two endpoints of a finite self-avoiding walk (SAW) in 2D, and indeed,
a broader class of curves.  Using exact but nonrigorous Coulomb gas
methods, they found that the distribution of winding angle approaches a
Gaussian and they explicitly computed the variance.  When the
endpoints of the walk are distance $L$ apart, the winding variance is
$\sim(8/g) \log L$, where $g$ is a model-dependent
parameter which is $3/2$ for SAW.
The winding angle at a single
endpoint (relative to the global average direction of the curve ---
see below for a precise definition) is a
Gaussian with variance $(4/g) \log L$ \cite{DS}.  We found
experimentally that the variance in the winding at typical (random) points
along the curve was only $1/4$ as large as the variance in the winding
at the endpoints.  More generally, when $k$ strands of the curve come
together at a point, the winding angle variance is $1/k^2$ as large
as at the endpoints;
Eq.~\eqref{eq:wind} below generalizes this further.

\textit{Remark:}
After our initial experiments we learned that the $4/(g k^2) \log L$
formula is also contained in unpublished notes of Duplantier.
However, to our knowledge the winding at typical points or points
where $k$ strands come together is not mentioned in the literature,
except in the case of loop-erased random walk (LERW), where we have
identified a minor oversight in the calculations that resulted in
incorrect values being reported.  Our experiments can be seen as a
test of Duplantier's Coulomb gas predictions.  We also report on other
random fractal curves, for which the Coulomb gas methods do not apply.

\sect{Fully-Packed Loop Model}
The main object of our study is the 2D Fortuin-Kasteleyn (FK) random
cluster model \cite{FK,Wu} at criticality, specifically the contours of
the clusters. 
The FK model is like bond percolation with edge probability $p$,
but there is another parameter $q$,
and $\Pr[\text{configuration}] \propto
p^{\text{\# bonds}} (1-p)^{\text{\# missing bonds}} q^{\text{\# clusters}}$.
For each $q$, there is a critical $p$ above which the system percolates.
When $p=p_{\text{critical}}$,
the contours of the clusters form a
system of loops called the fully-packed loop model
(FPL) \cite{BKW:FPL,nienhuis:FPL,dennijs:FPL}.
See Figure 1, left panels.
\begin{figure}[tphb]
\begin{center}
\epsfig{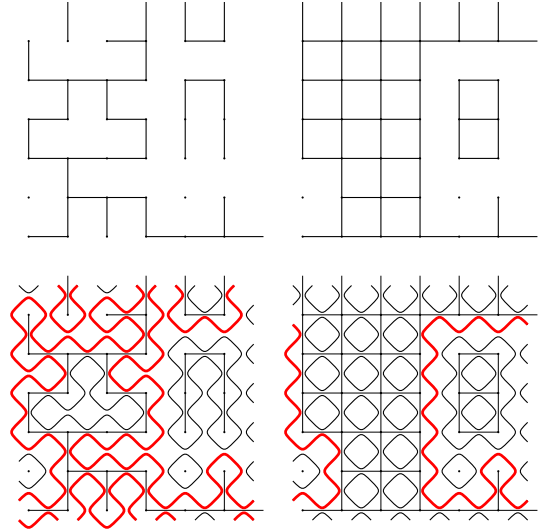}
\end{center}
\caption{
The contours and external perimeter of critical FK
clusters.  The upper left panel shows a FK random cluster
configuration with $q=3$ at criticality.
On the upper right panel we closed off the narrow
passageways of this FK configuration by connecting adjacent pairs
of vertices that belong to the same connected component.  In the lower
panels we show the fully packed loop configurations that come from the
above bond configurations.  The loops on the left traverse the hull or
perimeter of the clusters.  The loops on the right traverse the
external perimeter of the clusters.  The longest perimeter loop and
longest external perimeter loop are shown in bold.}
\label{fig:contours}
\end{figure}
A loop configuration occurs with probability
proportional to $n^{\text{\# loops}}$ where
$n=\sqrt{q}$ \cite{BKW:FPL,nienhuis:FPL,dennijs:FPL}.

\sect{Contours and External Perimeter}
In addition to the perimeter (or ``hull'') of a
cluster, the ``external perimeter'' has also been studied
(see Figure 1, right panels).  Grossman
and Aharony \cite{GA} found experimentally that by closing off narrow
passageways on the hull of percolation clusters, the fractal dimension
drops from $7/4$ \cite{SD} to $4/3$, and that furthermore the precise
definition of ``narrow'' had little or no effect on the fractal
dimension $4/3$.  See \cite{ADA} for an explanation of this phenomenon
in terms of path crossing exponents.  The external perimeter of
percolation clusters is believed to be essentially the self-avoiding
walk \cite{GA,CJMS,DS:SAW,ADA},
\begin{table*}[t]
\begin{tabular}{|c|c|c|c|c|c|c|c|c|c|@{$\,\,$}l||c|c|c|c|c|c|@{$\,\,$}l|}
\hline
\multicolumn{11}{|c||}{perimeter of FK clusters}& \multicolumn{7}{|c|}{external perimeter of FK clusters}\\
\hline
$q$ & $n$  & $g$  & $\kappa_1$ & $\kappa_2$ & $\kappa_3$ & $\kappa_4$ & $\kappa_5$ & $\kappa_6$ & $D_f$ & related model      &
$q$        & $g$  & $\kappa_1$ & $\kappa_2$ & $\kappa_3$ & $D_f$ & related model      \\
\hline
$0$ &	  
      $0$ & $1/2$ & $8$        & $2$ & & $1/2$ & & $2/9$  & $2$   & \begin{tabular}{@{}l@{}}spanning tree\\[-4pt]dense SAW\end{tabular} &
     $0$ & $2$   & $2$        & $1/2$      & $2/9$      & $5/4$ & LERW \\ 
$1$ &	  
      $1$ & $2/3$ & $6$        & $3/2$ & $2/3$ & & $6/25$ &      & $7/4$ & \begin{tabular}{@{}l@{}}polymers at $\Theta$ point\\[-4pt] square ice\end{tabular} &
      $1$ & $3/2$ & $8/3$      & $2/3$      &            & $4/3$ & \begin{tabular}{@{}l@{}}self avoiding walk\\[-4pt] Brownian frontier\end{tabular}\\
$2$ &	  
$\sqrt{2}$& $3/4$ & $16/3$     & $4/3$   &  &  &  &      & $5/3$ &              &
      $2$ & $4/3$ & $3$        & $3/4$      &            & $11/8$&              \\
$\!\frac{3+\sqrt{5}}{2}\!$ &
$\!\frac{1+\sqrt{5}}{2}\!$
          & $4/5$ & $5$        & $5/4$   &  &  &  &      & $13/8$&              &
$\!\frac{3+\sqrt{5}}{2}\!$
          & $5/4$ & $16/5$     &            &            & $7/5$ &              \\
$3$ &	  
$\sqrt{3}$& $5/6$ & $24/5$     & $6/5$   &  &  &  &      & $8/5$ &              &
$3$       & $6/5$ & $10/3$     &            &            &$17/12$&              \\
$4$ &	  
      $2$ & $1$   & $4$        & $1$     &  &  &  &      & $3/2$ & 2 perfect matchings&
$4$       & $1$   & $4$        &            &            & $3/2$ &              \\
\hline
\end{tabular}

\begin{tabular}{lcl}
$q$: cluster fugacity in FK random cluster model                   &\ \ &
     $n^2 = q$ \cite{BKW:FPL,nienhuis:FPL,dennijs:FPL} (perimeter only, not external perimeter)  \\[-1pt]

$n$: loop fugacity in fully packed loop model                      &&
     $n = -2\cos(\pi g)$      \cite{nienhuis:DL,KGN}      \\[-1pt]

$g$: coupling constant of associated Coulomb gas                   &&
     $\kappa=\kappa_1=4/g$    \cite{DS}                   \\[-1pt]

$\kappa_k$: winding angle variance when $k$ curves meet at a point &&
     $g_{\text{ext.\ perimeter}} = 1/g_{\text{perimeter}}$ \cite{D:EP}                                 \\[-1pt]

$D_f$: fractal dimension of curves                                 &&
     $D_f = 1+1/(2g) = 1+\kappa/8$ \cite{KGN,RS} 
\end{tabular}
\caption{Summary of $\kappa_k$ measurements for the perimeter and external
perimeter of FK clusters.
The table goes up to $q=4$, beyond which the FK model has a first order
phase transition without large loops \cite{Baxter,KT}.
The values of the fractal dimension $D_f$ are summarized from
\cite{nienhuis,SD,GA,ADA,majumdar,kondev-henley,kenyon:5/4,LSW:bf,D:EP,beffara:6},
and $\kappa_1$ comes from \cite{DS}.
The values for $\kappa_2$
for the perimeter and external perimeter were measured as described below.
$\kappa_3$ for LERW was measured by looking at the ``triple points'' of
uniformly random spanning trees -- points where three strands come together.
$\kappa_4$ and $\kappa_6$ for the spanning tree perimeter are
$\kappa_2$ and $\kappa_3$ for LERW respectively.  For $q=1$,
$\kappa_3$ is $\kappa_2$ of the external perimeter \cite{ADA,S:pc}, and
likewise $\kappa_5=\kappa_4'$ (defined below).  The measurements
are consistent with the hypothesis that $\kappa_k=\kappa_1/k^2$, a
formula that also appears in unpublished notes of Duplantier
\cite{duplantier:un}.
}
\label{tbl:kappa}
\end{table*}
and the external perimeters of FK clusters for other values of
$q$ are also interesting.  Therefore in addition to studying the hulls
of the FK clusters, we also studied their external perimeters by
closing off passageways of lattice spacing 1 and looking at the hulls
of the resulting clusters.

\sect{Related Models}
The perimeters and external perimeters of FK clusters are closely
related to a variety of models, most notably the stochastic Loewner
evolution (SLE) process introduced by Schramm \cite{S}.
In a discretized version of SLE, a
curve in the
plane grows as follows: the portion of the plane not in the curve is
conformally mapped to the half plane, with the tip of the curve mapped
to the origin.  A small random cut is then made starting at the origin, where
the slope of the cut is controlled by a parameter $\kappa$.  The
original curve gets extended by the pre-image
of this small cut, and
the process repeats.  The variance in the winding angle
at the endpoint of SLE$_\kappa$ is $\kappa \log L$ \cite{S}.

The SLE process describes the limiting behavior of a variety of
statistical mechanical models in 2D.  Schramm proved that SLE$_2$
gives the scaling limit of loop-erased random walk (LERW), provided
that LERW has a conformally invariant scaling limit; recently Lawler,
Schramm, and Werner \cite{LSW:2-8} proved this without assumptions.
Smirnov \cite{Smirnov} proved that critical site percolation in the
triangular lattice converges to SLE$_6$.  Lawler, Schramm, and Werner
\cite{LSW:8/3} proved that the frontier of Brownian motion (with
suitable boundary conditions) converges to SLE$_{8/3}$.  There are
good theoretical \cite{LSW:SAW} and experimental \cite{Kennedy}
reasons to believe that SLE$_{8/3}$ also describes the self-avoiding
walk (SAW).  Calculations by Kenyon and Schramm \cite{Kenyon-Schramm}
suggest that SLE$_4$ describes the loops arising from superimposing two
domino tilings.
Rohde and Schramm \cite{RS} proved
that the fractal dimension $D_f$ of SLE$_\kappa$ is at most $1+\kappa/8$ when $\kappa\leq 8$, and their calculations suggest $D_f=1+\kappa/8$.
See also \cite{LSW:bf,LSW:B1,LSW:B2,LSW:B3,Smirnov-Werner,Schramm,beffara:6}
for further results on SLE.
Schramm conjectured that
the contours of FK clusters at
criticality have the same local properties as
SLE$_\kappa$, where $\kappa$ depends on $q$.

\sect{Winding Angle Function}
We study the perimeter and external perimeter of
FK clusters by looking at the winding angle function.
Given a loop or a path in the plane or on a torus, we define the
winding angle function $w()$, a function of the edges, as follows.  We
pick an arbitrary starting edge $e$ on the loop or path and an arbitrary
value for the winding function $w(e)$ at that edge.  The
winding at a neighboring edge $e'$ is defined by
$w(e')=w(e)+$ the turning angle from $e$ to $e'$ measured in
radians.  This definition applies to paths or noncontractable
loops, i.e.\ loops that wind around the torus.  If a loop is
contractable to a point, this would yield a multivalued winding
function, so we adjust the definition of $w(e')-w(e)$ by
$2\pi/(\text{length of loop})$ to get a single-valued winding
function.  This specifies the winding angle function up to a global
additive constant; we choose the value of this global constant to
make the average winding angle of the edges on the loop or path $0$.

When $k$ strands of the perimeter or external perimeter converge on a point,
the winding angle variance should scale as $\kappa_k \log L$.
Table~\ref{tbl:kappa} summarizes our measurements of $\kappa_k$,
suggesting $\kappa_k=\kappa_1/k^2$
(see also \cite{duplantier:un,duplantier-binder:winding}).

\sect{Remarks on LERW}
\textit{Remarks on LERW:\/}
Our simulation values for $\kappa_2$ and $\kappa_3$ for
LERW disagree with the values previously reported by Kenyon \cite{K2}
by a factor of $4$ and $9$ respectively.  These calculations used the
Temperley \cite{T} correspondence between spanning trees and dimer
systems, and Kenyon correctly and rigorously computed the variance in
the height function of the associated dimer system when there were 1,
2, or 3 paths approaching a point.
The height function of the dimer system is related to the
winding angle for the paths: when there are $k$ paths, each winding
changes the dimer height function by $4 k$.  The factor of $k$ was
omitted, leading to the factor of $k^2$ discrepancy in the winding
angle variance.
\vspace*{3pt}

\sect{Windings at Pinch Points}
The longest contour of an FK configuration is likely to hit itself
many times (which is why the perimeter and external perimeter are
different); the places where the contour hits itself are called pinch
points.  For the longest contour we identified the pinch point giving
rise to the longest pinch.  At this point there are four strands that
travel a distance on the order of the box length $L$, suggesting that
the winding angle variance at this point should grow as $\kappa_4 \log
L$.  However, as noted by Schramm \cite{S:pc}, the pinch point with
the longest pinch is an atypical pinch point because there are two adjacent
strands are conditioned not to hit each other --- if they did hit each
other, then this would create a longer pinch.  Thus the winding angle
variance at the longest pinch point is governed by a different
constant $\kappa_4'$, and grows as $\kappa_4' \log L$.

In general let $\kappa_k'$ be the winding angle variance coefficient
when there are $k$ strands meeting at a point and two adjacent
strands do not hit each other (when $k=2$, the left side of one strand
may hit the right side of the other strand, but not vice versa).  When $q=4$,
the strands do not hit each other anyway \cite{RS}, so
$\kappa_k'=\kappa_k$.  When $q=1$ and $k$ strands meet
at a point, conditioning two adjacent strands not to touch has the same
effect as adding an extra strand: $\kappa_k'=\kappa_{k+1}$ \cite{ADA,S:pc}.
For other values of $q$ it is plausible that requiring two of the
strands not to hit each other has the effect of adding some fractional
number of strands $f(q)$ between the strands required not to hit each
other; similar phenomena have been observed elsewhere.
When two strands meet at a point that happens to be on the external
perimeter, the right side of one strand does not hit the left side
of the other strand.
Thus $\kappa_2'$ for the perimeter is $\kappa_2$ for the external perimeter,
which (by $\kappa_2=\kappa_1/4=1/g$ \cite{DS} and
$g_{\text{ext.\ perimeter}} = 1/g_{\text{perimeter}}$ \cite{D:EP})
in turn is $1/\kappa_2$ (for the perimeter), giving
\begin{align*}
\kappa_1/(2+f(q))^2 &= 4/\kappa_1\\
 f(q) &= \kappa_1/2 - 2 \\
 \kappa_k' &= \kappa_1/(k+\kappa_1/2-2)^2 .
\end{align*}
For example, when $q=0$ this predicts $\kappa_4'=2/9$.  Indeed, the
largest pinch point for SLE$_8$ corresponds to a triple point of the
spanning tree, for which we already have the value $2/9$.
For other values of $q$, our measured values of $\kappa_4'$ appear
to be consistent with this formula.

More generally, when $k$ strands meet at a point, and
$j$ adjacent pairs do not hit each other, we expect the winding angle
variance to grow like
\begin{equation}
\label{eq:wind}
  \frac{\kappa_1}{(k+ j \max(0,\kappa_1/2-2))^2} \log L.
\tag{\ensuremath{\star}}
\end{equation}

\sect{Measuring the $\kappa$'s}
To measure the $\kappa$'s,
for a given value of $q$, for each of several system sizes
 (side length
$L$ a power of two multiple of 4, 5, 6, or 7, starting with
$L=4\times 2^3,5\times 2^3,6\times 2^3,7\times 2^3,4\times 2^4,\ldots$),
 we generated 10000 random
FK configurations using the methods
described in \cite{PW1,PW2}.
In each one we identified the
longest contour (and also the longest outer contour) and computed the
winding angle function as defined above.  For $\kappa_2$,
the square of the winding at a random single edge on the loop is an
estimator of the winding angle variance of the loop, but a more
efficient estimator is the average square of the winding angle of
edges on the loop.  For $\kappa_4'$, we measured the square
of the winding at the pinch point of the longest pinch of the longest
contour.  The data for the longest
contour when $q=1$ (percolation) is shown in
Figure~\ref{fig:q=1,kappa2}, whose caption explains how we estimated
$\kappa_2$, $\kappa_2'$, and $\kappa_4'$.
Tables~\ref{tbl:kappa-est} and \ref{tbl:tree} summarize our
estimates.

\begin{figure}[tphb]
\begin{center}
\epsfig{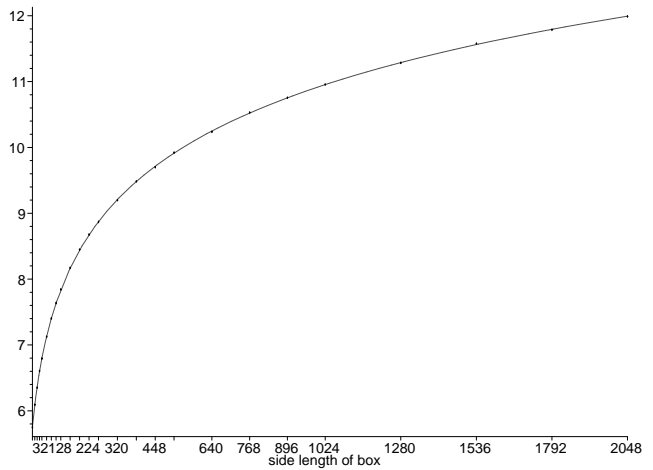}
\caption{
Winding angle variance for the largest contour when $q=1$ as a
function of the side length $L$ of the box.  Error bars on our
estimates of the winding angle variance are shown, but are quite short
and appear as points.  A curve of the form $\kappa_2 \log L + a$ was
least-squares fitted to this data and plotted here.  The 95\%
confidence intervals ($\pm1.96$ standard deviations) for the
parameters are $\kappa_2=1.5002\pm 0.0023$ and $a=0.55494\pm 0.013$,
consistent with $\kappa_2=3/2$.  The fit has a $\chi^2$ statistic of
$23.65$ with $23$ degrees of freedom for a $p$-value of $0.42$, so
the fit passes the $\chi^2$ test.  In this case the fit is good all the
way down to $L_{\min}=32$.  In some cases the fitted curve lies outside the
95\% confidence interval at $L_{\min}$, indicating
corrections to scaling,
and in these cases we increased $L_{\min}$.
These data are summarized in the first line of Table~\ref{tbl:kappa-est}.}
\label{fig:q=1,kappa2}
\end{center}
\end{figure}

We also conducted measurements for the minimum spanning tree (MST)
with random edge weights.  The paths of the MST are smoother and
less windy than those of the UST (see Table~\ref{tbl:tree}).
For MST it is unlikely that $D_f = 1+\kappa_1/8$, so the MST path is not
described by SLE.

In conclusion, Eq.~\eqref{eq:wind}, which generalizes Duplantier's
winding angle formula, is supported by both experiments and heuristic arguments.
It would be interesting to see if Eq.~\eqref{eq:wind} holds for SLE$_\kappa$.

\vspace*{1.2pt}
\sect{Acknowledgements}
\textit{Acknowledgements:\/}
We thank Oded Schramm and Jan\'e Kondev for valuable discussions.

\pagebreak

\newcommand{\rest}{\begin{tabular}{@{}c@{}}nearby\\[-4pt]rational\end{tabular} & $L$'s & \begin{tabular}{c}$\chi^2$-test\\[-4pt]$p$-value\end{tabular}}
\begin{table}[tphb]
\begin{center}
\begin{tabular}{|c|c|c|c|c|c|c|}
\hline
 $q$ & $\kappa_2$ & \rest \\ \hline
1 & $1.500\pm.002$ &   3/2& 32--2048 & $ 0.42$ \\
2 & $1.333\pm.003$ &   4/3& 32--1280 & $ 0.77$ \\
$\!\frac{3+\!\sqrt{5}}{2}\!$
  & $1.252\pm.003$ &   5/4&  32--896 & $ 0.036$ \\
3 & $1.204\pm.004$ &   6/5&  32--896 & $ 0.59$ \\
4 & $1.078\pm.007$ &    1?&  32--768 & $ 0.72$ \\
\hline
\hline $q$ & $\kappa_2'$ & \rest \\ \hline
1 & $0.666\pm.002$ &   2/3& 80--2048 & $ 0.71$ \\
2 & $0.747\pm.002$ &   3/4& 32--1280 & $ 0.67$ \\
$\!\frac{3+\!\sqrt{5}}{2}\!$
  & $0.779\pm.003$ &  4/5?&  32--896 & $ 0.66$ \\
3 & $0.795\pm.005$ &    ??&  32--896 & $ 0.019$ \\
4 & $0.800\pm.008$ &    ??&  32--768 & $ 0.75$ \\
\hline
\hline $q$ & $\kappa_4'$ & \rest \\ \hline
1 & $0.239\pm.007$ &  6/25& 32--2048 & $ 0.84$ \\
2 & $0.247\pm.008$ &12/49?& 32--1280 & $ 0.27$ \\
$\!\frac{3+\!\sqrt{5}}{2}\!$
  & $0.243\pm.009$ &20/81?&  32--896 & $ 0.31$ \\
3 & $0.243\pm.009$ &30/121?&  32--896 & $ 0.71$ \\
4 & $0.267\pm.010$ & 1/4??&  32--768 & $ 0.86$ \\
\hline

\end{tabular}
\caption{
Winding angle variance coefficient for the longest loop ($\kappa_2$),
longest loop of the external perimeter ($\kappa_2'$), and
largest pinch of the longest loop ($\kappa_4'$).
When $q=4$, we expect that log-corrections
\cite{salas-sokal:q4,aharony-asikainen:q4}
affect the measured $\kappa_2'$ and $\kappa_4'$.}
\label{tbl:kappa-est}
\end{center}
\end{table}
\begin{table}[tphb]
\begin{center}
\begin{tabular}{|c|c|c|c|c|}
\cline{2-5}  \multicolumn{1}{c|}{\begin{tabular}{@{}c@{}}UST path\\[-12pt]\phantom{MST path}\\[-4pt](LERW)\end{tabular}} & parameter & \rest \\ \hline
$\kappa_1$ & $2.000\pm.002$ &     2& 32--1792 & $ 0.024$ \\
$\kappa_2$ & $0.510\pm.003$ &   1/2& 48--1792 & $ 0.60$ \\
$\kappa_3$ & $0.235\pm.010$ &   2/9& 40--1792 & $ 0.17$ \\
$\kappa_3$ & $0.229\pm.008$ &   2/9& 32--1792 & $ 0.34$ \\
$D_f$ & $1.252\pm.001$ &   5/4& 32--1792 & $ 0.83$ \\
\hline
\end{tabular}
\vspace{\doublerulesep}
\begin{tabular}{|c|c|c|c|c|}
\cline{2-5} \multicolumn{1}{c|}{MST path} & parameter & \rest \\ \hline
&&&\phantom{32--1024}&\\[-12pt]
$\kappa_1$ & $1.883\pm.002$ &     ?& 32--2048 & $ 0.86$ \\
$\kappa_2$ & $0.442\pm.002$ &     ?& 32--2048 & $ 0.055$ \\
$\kappa_3$ & $0.199\pm.007$ &     ?& 32--2048 & $ 0.61$ \\
$\kappa_3$ & $0.199\pm.007$ &     ?& 32--2048 & $ 0.93$ \\
$D_f$ & $1.218\pm.001$ &     ?& 32--2048 & $ 0.10$ \\
\hline
\end{tabular}

\caption{
Measurements of $\kappa_1$, $\kappa_2$, $\kappa_3$, and $D_f$ for the
paths in uniform spanning tree (UST) (i.e.\ LERW \cite{pemantle:tree})
and the minimum spanning tree (MST).  
The estimate of $\kappa_1$ comes from $\kappa_2$ of the spanning tree contour.
The first estimate of $\kappa_3$ comes from a triple point, the second estimate comes from the longest pinch ($\kappa_4'$) of the spanning tree contour.
}
\label{tbl:tree}
\end{center}
\end{table}

\bibliography{ww}

\begin{thebibliography}{45}
\expandafter\ifx\csname natexlab\endcsname\relax\def\natexlab#1{#1}\fi
\expandafter\ifx\csname bibnamefont\endcsname\relax
  \def\bibnamefont#1{#1}\fi
\expandafter\ifx\csname bibfnamefont\endcsname\relax
  \def\bibfnamefont#1{#1}\fi
\expandafter\ifx\csname citenamefont\endcsname\relax
  \def\citenamefont#1{#1}\fi
\expandafter\ifx\csname url\endcsname\relax
  \def\url#1{\texttt{#1}}\fi
\expandafter\ifx\csname urlprefix\endcsname\relax\def\urlprefix{URL }\fi
\providecommand{\bibinfo}[2]{#2}
\providecommand{\eprint}[2][]{\url{#2}}

\bibitem[{\citenamefont{Duplantier and Saleur}(1988)}]{DS}
\bibinfo{author}{\bibfnamefont{B.}~\bibnamefont{Duplantier}} \bibnamefont{and}
  \bibinfo{author}{\bibfnamefont{H.}~\bibnamefont{Saleur}},
  \bibinfo{journal}{Phys. Rev. Lett.} \textbf{\bibinfo{volume}{60}},
  \bibinfo{pages}{2343} (\bibinfo{year}{1988}).

\bibitem[{\citenamefont{Fortuin and Kasteleyn}(1972)}]{FK}
\bibinfo{author}{\bibfnamefont{C.}~\bibnamefont{Fortuin}} \bibnamefont{and}
  \bibinfo{author}{\bibfnamefont{P.}~\bibnamefont{Kasteleyn}},
  \bibinfo{journal}{Physica} \textbf{\bibinfo{volume}{57}},
  \bibinfo{pages}{536} (\bibinfo{year}{1972}).

\bibitem[{\citenamefont{Wu}(1982)}]{Wu}
\bibinfo{author}{\bibfnamefont{F.~Y.} \bibnamefont{Wu}}, \bibinfo{journal}{Rev.
  Mod. Phys.} \textbf{\bibinfo{volume}{54}}, \bibinfo{pages}{235}
  (\bibinfo{year}{1982}).

\bibitem[{\citenamefont{Baxter et~al.}(1976)\citenamefont{Baxter, Kelland, and
  Wu}}]{BKW:FPL}
\bibinfo{author}{\bibfnamefont{R.~J.} \bibnamefont{Baxter}},
  \bibinfo{author}{\bibfnamefont{S.~B.} \bibnamefont{Kelland}},
  \bibnamefont{and} \bibinfo{author}{\bibfnamefont{F.~Y.} \bibnamefont{Wu}},
  \bibinfo{journal}{J. Phys.\ A} \textbf{\bibinfo{volume}{9}},
  \bibinfo{pages}{397} (\bibinfo{year}{1976}).

\bibitem[{\citenamefont{den Nijs}(1983)}]{dennijs:FPL}
\bibinfo{author}{\bibfnamefont{M.}~\bibnamefont{den Nijs}},
  \bibinfo{journal}{Phys. Rev. B} \textbf{\bibinfo{volume}{27}},
  \bibinfo{pages}{1674} (\bibinfo{year}{1983}).

\bibitem[{\citenamefont{Nienhuis}(1982)}]{nienhuis:FPL}
\bibinfo{author}{\bibfnamefont{B.}~\bibnamefont{Nienhuis}},
  \bibinfo{journal}{Phys. Rev. Lett.} \textbf{\bibinfo{volume}{49}},
  \bibinfo{pages}{1062} (\bibinfo{year}{1982}).

\bibitem[{\citenamefont{Grossman and Aharony}(1987)}]{GA}
\bibinfo{author}{\bibfnamefont{T.}~\bibnamefont{Grossman}} \bibnamefont{and}
  \bibinfo{author}{\bibfnamefont{A.}~\bibnamefont{Aharony}},
  \bibinfo{journal}{J. Phys.\ A} \textbf{\bibinfo{volume}{20}},
  \bibinfo{pages}{L1193} (\bibinfo{year}{1987}).

\bibitem[{\citenamefont{Saleur and Duplantier}(1987)}]{SD}
\bibinfo{author}{\bibfnamefont{H.}~\bibnamefont{Saleur}} \bibnamefont{and}
  \bibinfo{author}{\bibfnamefont{B.}~\bibnamefont{Duplantier}},
  \bibinfo{journal}{Phys. Rev. Lett.} \textbf{\bibinfo{volume}{58}},
  \bibinfo{pages}{2325} (\bibinfo{year}{1987}).

\bibitem[{\citenamefont{Aizenman et~al.}(1999)\citenamefont{Aizenman,
  Duplantier, and Aharony}}]{ADA}
\bibinfo{author}{\bibfnamefont{M.}~\bibnamefont{Aizenman}},
  \bibinfo{author}{\bibfnamefont{B.}~\bibnamefont{Duplantier}},
  \bibnamefont{and} \bibinfo{author}{\bibfnamefont{A.}~\bibnamefont{Aharony}},
  \bibinfo{journal}{Phys. Rev. Lett.} \textbf{\bibinfo{volume}{83}},
  \bibinfo{pages}{1359} (\bibinfo{year}{1999}), \eprint{cond-mat/9901018}.

\bibitem[{\citenamefont{Coniglio et~al.}(1987)\citenamefont{Coniglio, Jan,
  Majid, and Stanley}}]{CJMS}
\bibinfo{author}{\bibfnamefont{A.}~\bibnamefont{Coniglio}},
  \bibinfo{author}{\bibfnamefont{N.}~\bibnamefont{Jan}},
  \bibinfo{author}{\bibfnamefont{I.}~\bibnamefont{Majid}}, \bibnamefont{and}
  \bibinfo{author}{\bibfnamefont{H.~E.} \bibnamefont{Stanley}},
  \bibinfo{journal}{Phys. Rev. B} \textbf{\bibinfo{volume}{35}},
  \bibinfo{pages}{3617} (\bibinfo{year}{1987}).

\bibitem[{\citenamefont{Duplantier and Saleur}(1987)}]{DS:SAW}
\bibinfo{author}{\bibfnamefont{B.}~\bibnamefont{Duplantier}} \bibnamefont{and}
  \bibinfo{author}{\bibfnamefont{H.}~\bibnamefont{Saleur}},
  \bibinfo{journal}{Phys. Rev. Lett.} \textbf{\bibinfo{volume}{59}},
  \bibinfo{pages}{539} (\bibinfo{year}{1987}).

\bibitem[{\citenamefont{Kondev et~al.}(1996)\citenamefont{Kondev, de~Gier, and
  Nienhuis}}]{KGN}
\bibinfo{author}{\bibfnamefont{J.}~\bibnamefont{Kondev}},
  \bibinfo{author}{\bibfnamefont{J.}~\bibnamefont{de~Gier}}, \bibnamefont{and}
  \bibinfo{author}{\bibfnamefont{B.}~\bibnamefont{Nienhuis}},
  \bibinfo{journal}{J. Phys.\ A} \textbf{\bibinfo{volume}{29}},
  \bibinfo{pages}{6489} (\bibinfo{year}{1996}), \eprint{cond-mat/9603170}.

\bibitem[{\citenamefont{Nienhuis}(1987)}]{nienhuis:DL}
\bibinfo{author}{\bibfnamefont{B.}~\bibnamefont{Nienhuis}}, in
  \emph{\bibinfo{booktitle}{Phase Transitions and Critical Phenomena \#11}},
  edited by \bibinfo{editor}{\bibnamefont{{Domb \& Lebowitz}}}
  (\bibinfo{year}{1987}), pp. \bibinfo{pages}{1--53}.

\bibitem[{\citenamefont{Duplantier}(2000)}]{D:EP}
\bibinfo{author}{\bibfnamefont{B.}~\bibnamefont{Duplantier}},
  \bibinfo{journal}{Phys. Rev. Lett.} \textbf{\bibinfo{volume}{84}},
  \bibinfo{pages}{1363} (\bibinfo{year}{2000}).

\bibitem[{\citenamefont{Rohde and Schramm}(2001)}]{RS}
\bibinfo{author}{\bibfnamefont{S.}~\bibnamefont{Rohde}} \bibnamefont{and}
  \bibinfo{author}{\bibfnamefont{O.}~\bibnamefont{Schramm}}
  (\bibinfo{year}{2001}), \eprint{math.PR/0106036}.

\bibitem[{\citenamefont{Baxter}(1973)}]{Baxter}
\bibinfo{author}{\bibfnamefont{R.~J.} \bibnamefont{Baxter}},
  \bibinfo{journal}{J. Phys.\ C} \textbf{\bibinfo{volume}{6}},
  \bibinfo{pages}{L445} (\bibinfo{year}{1973}).

\bibitem[{\citenamefont{Kosterlitz and Thouless}(1973)}]{KT}
\bibinfo{author}{\bibfnamefont{J.}~\bibnamefont{Kosterlitz}} \bibnamefont{and}
  \bibinfo{author}{\bibfnamefont{D.}~\bibnamefont{Thouless}},
  \bibinfo{journal}{J. Phys.\ C} \textbf{\bibinfo{volume}{6}},
  \bibinfo{pages}{1181} (\bibinfo{year}{1973}).

\bibitem[{\citenamefont{Beffara}(2002)}]{beffara:6}
\bibinfo{author}{\bibfnamefont{V.}~\bibnamefont{Beffara}}
  (\bibinfo{year}{2002}), \eprint{math.PR/0204208}.

\bibitem[{\citenamefont{Kenyon}(2000{\natexlab{a}})}]{kenyon:5/4}
\bibinfo{author}{\bibfnamefont{R.}~\bibnamefont{Kenyon}},
  \bibinfo{journal}{Acta Math.} \textbf{\bibinfo{volume}{185}},
  \bibinfo{pages}{239} (\bibinfo{year}{2000}{\natexlab{a}}).

\bibitem[{\citenamefont{Lawler et~al.}(2001{\natexlab{a}})\citenamefont{Lawler,
  Schramm, and Werner}}]{LSW:bf}
\bibinfo{author}{\bibfnamefont{G.~F.} \bibnamefont{Lawler}},
  \bibinfo{author}{\bibfnamefont{O.}~\bibnamefont{Schramm}}, \bibnamefont{and}
  \bibinfo{author}{\bibfnamefont{W.}~\bibnamefont{Werner}},
  \bibinfo{journal}{Math.\ Res.\ Lett.} \textbf{\bibinfo{volume}{8}},
  \bibinfo{pages}{401} (\bibinfo{year}{2001}{\natexlab{a}}),
  \eprint{math.PR/0010165}.

\bibitem[{\citenamefont{Majumdar}(1992)}]{majumdar}
\bibinfo{author}{\bibfnamefont{S.~N.} \bibnamefont{Majumdar}},
  \bibinfo{journal}{Phys. Rev. Lett.} \textbf{\bibinfo{volume}{68}},
  \bibinfo{pages}{2329} (\bibinfo{year}{1992}).

\bibitem[{\citenamefont{Nienhuis}(1984)}]{nienhuis}
\bibinfo{author}{\bibfnamefont{B.}~\bibnamefont{Nienhuis}},
  \bibinfo{journal}{J. Stat. Phys.} \textbf{\bibinfo{volume}{34}},
  \bibinfo{pages}{731} (\bibinfo{year}{1984}).

\bibitem[{\citenamefont{Kondev and Henley}(1995)}]{kondev-henley}
\bibinfo{author}{\bibfnamefont{J.}~\bibnamefont{Kondev}} \bibnamefont{and}
  \bibinfo{author}{\bibfnamefont{C.~L.} \bibnamefont{Henley}},
  \bibinfo{journal}{Phys. Rev. Lett.} \textbf{\bibinfo{volume}{74}},
  \bibinfo{pages}{4580} (\bibinfo{year}{1995}).

\bibitem[{\citenamefont{Schramm}(2002)}]{S:pc}
\bibinfo{author}{\bibfnamefont{O.}~\bibnamefont{Schramm}}
  (\bibinfo{year}{2002}), \bibinfo{note}{personal communication}.

\bibitem[{\citenamefont{Duplantier}()}]{duplantier:un}
\bibinfo{author}{\bibfnamefont{B.}~\bibnamefont{Duplantier}},
  \bibinfo{note}{unpublished notes}.

\bibitem[{\citenamefont{Schramm}(2000)}]{S}
\bibinfo{author}{\bibfnamefont{O.}~\bibnamefont{Schramm}},
  \bibinfo{journal}{Israel J. Math.\xspace} \textbf{\bibinfo{volume}{118}},
  \bibinfo{pages}{221} (\bibinfo{year}{2000}).

\bibitem[{\citenamefont{Lawler et~al.}(2001{\natexlab{b}})\citenamefont{Lawler,
  Schramm, and Werner}}]{LSW:2-8}
\bibinfo{author}{\bibfnamefont{G.~F.} \bibnamefont{Lawler}},
  \bibinfo{author}{\bibfnamefont{O.}~\bibnamefont{Schramm}}, \bibnamefont{and}
  \bibinfo{author}{\bibfnamefont{W.}~\bibnamefont{Werner}}
  (\bibinfo{year}{2001}{\natexlab{b}}), \eprint{math.PR/0112234}.

\bibitem[{\citenamefont{Smirnov}(2001)}]{Smirnov}
\bibinfo{author}{\bibfnamefont{S.}~\bibnamefont{Smirnov}}
  (\bibinfo{year}{2001}),
  \urlprefix\url{http://www.math.kth.se/~stas/papers/percol.ps}.

\bibitem[{\citenamefont{Lawler et~al.}(2002{\natexlab{a}})\citenamefont{Lawler,
  Schramm, and Werner}}]{LSW:8/3}
\bibinfo{author}{\bibfnamefont{G.~F.} \bibnamefont{Lawler}},
  \bibinfo{author}{\bibfnamefont{O.}~\bibnamefont{Schramm}}, \bibnamefont{and}
  \bibinfo{author}{\bibfnamefont{W.}~\bibnamefont{Werner}}
  (\bibinfo{year}{2002}{\natexlab{a}}), \bibinfo{note}{in preparation}.

\bibitem[{\citenamefont{Lawler et~al.}(2002{\natexlab{b}})\citenamefont{Lawler,
  Schramm, and Werner}}]{LSW:SAW}
\bibinfo{author}{\bibfnamefont{G.~F.} \bibnamefont{Lawler}},
  \bibinfo{author}{\bibfnamefont{O.}~\bibnamefont{Schramm}}, \bibnamefont{and}
  \bibinfo{author}{\bibfnamefont{W.}~\bibnamefont{Werner}}
  (\bibinfo{year}{2002}{\natexlab{b}}), \eprint{math.PR/0204277}.

\bibitem[{\citenamefont{Kennedy}(2002)}]{Kennedy}
\bibinfo{author}{\bibfnamefont{T.}~\bibnamefont{Kennedy}},
  \bibinfo{journal}{Phys. Rev. Lett.} \textbf{\bibinfo{volume}{88}},
  \bibinfo{pages}{130601} (\bibinfo{year}{2002}).

\bibitem[{\citenamefont{Kenyon and Schramm}(1998)}]{Kenyon-Schramm}
\bibinfo{author}{\bibfnamefont{R.}~\bibnamefont{Kenyon}} \bibnamefont{and}
  \bibinfo{author}{\bibfnamefont{O.}~\bibnamefont{Schramm}}
  (\bibinfo{year}{1998}), \bibinfo{note}{unpublished}.

\bibitem[{\citenamefont{Lawler et~al.}(2001{\natexlab{c}})\citenamefont{Lawler,
  Schramm, and Werner}}]{LSW:B1}
\bibinfo{author}{\bibfnamefont{G.~F.} \bibnamefont{Lawler}},
  \bibinfo{author}{\bibfnamefont{O.}~\bibnamefont{Schramm}}, \bibnamefont{and}
  \bibinfo{author}{\bibfnamefont{W.}~\bibnamefont{Werner}},
  \bibinfo{journal}{Acta Math.} \textbf{\bibinfo{volume}{187}},
  \bibinfo{pages}{237} (\bibinfo{year}{2001}{\natexlab{c}}),
  \eprint{math.PR/9911084}.

\bibitem[{\citenamefont{Lawler et~al.}(2001{\natexlab{d}})\citenamefont{Lawler,
  Schramm, and Werner}}]{LSW:B2}
\bibinfo{author}{\bibfnamefont{G.~F.} \bibnamefont{Lawler}},
  \bibinfo{author}{\bibfnamefont{O.}~\bibnamefont{Schramm}}, \bibnamefont{and}
  \bibinfo{author}{\bibfnamefont{W.}~\bibnamefont{Werner}},
  \bibinfo{journal}{Acta Math.} \textbf{\bibinfo{volume}{187}},
  \bibinfo{pages}{275} (\bibinfo{year}{2001}{\natexlab{d}}),
  \eprint{math.PR/0003156}.

\bibitem[{\citenamefont{Lawler et~al.}(2002{\natexlab{c}})\citenamefont{Lawler,
  Schramm, and Werner}}]{LSW:B3}
\bibinfo{author}{\bibfnamefont{G.~F.} \bibnamefont{Lawler}},
  \bibinfo{author}{\bibfnamefont{O.}~\bibnamefont{Schramm}}, \bibnamefont{and}
  \bibinfo{author}{\bibfnamefont{W.}~\bibnamefont{Werner}},
  \bibinfo{journal}{Ann.\ I. H. Poincar\'e -- PR}
  \textbf{\bibinfo{volume}{38}}, \bibinfo{pages}{109}
  (\bibinfo{year}{2002}{\natexlab{c}}), \eprint{math.PR/0005294}.

\bibitem[{\citenamefont{Schramm}(2001)}]{Schramm}
\bibinfo{author}{\bibfnamefont{O.}~\bibnamefont{Schramm}},
  \bibinfo{journal}{Electron.\ Comm.\ Probab.\xspace}
  \textbf{\bibinfo{volume}{6}}, \bibinfo{pages}{115} (\bibinfo{year}{2001}).

\bibitem[{\citenamefont{Smirnov and Werner}(2001)}]{Smirnov-Werner}
\bibinfo{author}{\bibfnamefont{S.}~\bibnamefont{Smirnov}} \bibnamefont{and}
  \bibinfo{author}{\bibfnamefont{W.}~\bibnamefont{Werner}},
  \bibinfo{journal}{Math\ Res.\ Let.} \textbf{\bibinfo{volume}{8}},
  \bibinfo{pages}{729} (\bibinfo{year}{2001}).

\bibitem[{\citenamefont{Duplantier and
  Binder}(2002)}]{duplantier-binder:winding}
\bibinfo{author}{\bibfnamefont{B.}~\bibnamefont{Duplantier}} \bibnamefont{and}
  \bibinfo{author}{\bibfnamefont{I.~A.} \bibnamefont{Binder}}
  (\bibinfo{year}{2002}), \eprint{cond-mat/0208045}.

\bibitem[{\citenamefont{Kenyon}(2000{\natexlab{b}})}]{K2}
\bibinfo{author}{\bibfnamefont{R.}~\bibnamefont{Kenyon}}, \bibinfo{journal}{J.
  Math. Phys.} \textbf{\bibinfo{volume}{41}}, \bibinfo{pages}{1338}
  (\bibinfo{year}{2000}{\natexlab{b}}).

\bibitem[{\citenamefont{Temperley}(1974)}]{T}
\bibinfo{author}{\bibfnamefont{H.~N.~V.} \bibnamefont{Temperley}}, in
  \emph{\bibinfo{booktitle}{London Math.\ Soc.\ Lecture Notes Series \#13}}
  (\bibinfo{year}{1974}), pp. \bibinfo{pages}{202--204}.

\bibitem[{\citenamefont{Propp and Wilson}(1996)}]{PW1}
\bibinfo{author}{\bibfnamefont{J.G.}~\bibnamefont{Propp}} \bibnamefont{and}
  \bibinfo{author}{\bibfnamefont{D.B.}~\bibnamefont{Wilson}},
  \bibinfo{journal}{Ran.\ St.\ Alg.\xspace} \textbf{\bibinfo{volume}{9}},
  \bibinfo{pages}{223} (\bibinfo{year}{1996}).

\bibitem[{\citenamefont{Propp and Wilson}(1998)}]{PW2}
\bibinfo{author}{\bibfnamefont{J.~G.} \bibnamefont{Propp}} \bibnamefont{and}
  \bibinfo{author}{\bibfnamefont{D.~B.} \bibnamefont{Wilson}},
  \bibinfo{journal}{J. Alg.\xspace} \textbf{\bibinfo{volume}{27}},
  \bibinfo{pages}{170} (\bibinfo{year}{1998}).

\bibitem[{\citenamefont{Salas and Sokal}(1997)}]{salas-sokal:q4}
\bibinfo{author}{\bibfnamefont{J.}~\bibnamefont{Salas}} \bibnamefont{and}
  \bibinfo{author}{\bibfnamefont{A.~D.} \bibnamefont{Sokal}},
  \bibinfo{journal}{J. Stat. Phys.} \textbf{\bibinfo{volume}{88}},
  \bibinfo{pages}{567} (\bibinfo{year}{1997}).

\bibitem[{\citenamefont{Aharony and Asikainen}(2002)}]{aharony-asikainen:q4}
\bibinfo{author}{\bibfnamefont{A.}~\bibnamefont{Aharony}} \bibnamefont{and}
  \bibinfo{author}{\bibfnamefont{J.}~\bibnamefont{Asikainen}}
  (\bibinfo{year}{2002}), \eprint{cond-mat/0206367}.

\bibitem[{\citenamefont{Pemantle}(1991)}]{pemantle:tree}
\bibinfo{author}{\bibfnamefont{R.}~\bibnamefont{Pemantle}},
  \bibinfo{journal}{Ann.\ Prob.\xspace} \textbf{\bibinfo{volume}{19}},
  \bibinfo{pages}{1559} (\bibinfo{year}{1991}).

\end{thebibliography}

\end{document}